\documentclass[%
 reprint,
 amsmath,amssymb,
 aps,
]{revtex4-2}

\usepackage{graphicx}
\usepackage{dcolumn}
\usepackage{bm}

\begin{document}
\preprint{APS/123-QED}

\title{Physics of Smart Matter: integrating active matter and control to gain insights into living systems}

\author{Herbert Levine$^{1}$ and Daniel I. Goldman$^{2}$}
\
 \address{$^1$ Departments of Physics and Bioengineering, Northeastern University, Boston, MA, USA}
\address{$^2$ School of Physics, Georgia Institute of Technology, Atlanta, Georgia 30332, USA}%

\email{$^1$h.levine@northeastern.edu}
\email{$^2$daniel.goldman@physics.gatech.edu}

\date{\today}

\begin{abstract}
We offer our opinion on the benefits of integration of insights from active matter physics with principles of regulatory interactions and control to develop a field we term ``smart matter". This field can provide insight into important principles in living systems as well as aid engineering of responsive, robust and functional collectives. 

\end{abstract}

\maketitle


Recent years have seen the dawning of a new vibrant subfield of physics, the physics of living systems. An excellent discussion of the history and promise of this research area was presented in a recent National Academy report~\cite{national2022pols}. The research agenda of this field encompasses systems ranging from biomolecules to ecosystems, from building models to building robots~\cite{aguilar2016review}, all in service of a quantitative understanding of systems whose behavior transcends what we have come to expect from experience with inanimate matter. Here we offer our opinion as to what might be for a physical scientist some of the necessary ingredients to consider a system to be living. We will argue that a path toward addressing such questions and characterizing these systems will be in developing ideas of ``smart matter" (which we will define below) that combine ideas of the exploding field of active matter with those of the less appreciated (by physicists) regulatory interactions (e.g. feedback control~\cite{aastrom2021feedback,cowan2014feedback}).\\

Our starting point is the idea of active matter~\cite{marchetti}, a field that many physicists have come to believe underlies the secrets of life. Active matter refers to physical organizations of interacting constituents that each have their own access to energy sources. These constituents can be living, as is the case in bacterial colonies~\cite{bacteria}, ant rafts~\cite{rafts} or bird flocks~\cite{flocks}, completely abiotic as in colloids propelled by catalyzed chemical reactions or motor driven robots~\cite{li2021programming,aguilar2018collective}, or ``in-between” as in the beautiful dynamical structures created in vitro by biopolymers activated by molecular motors~\cite{nedelec} or even biohybrid robots composed of soft materials and living cells~\cite{hybrid}. The study of active matter in the physics community took off with the seminal work of Ben-Jacob, Vicsek and collaborators~\cite{vicsek} who showed that these systems can self-organize in ways that circumvent many of the restrictions exhibited by ``normal matter”,  operating close to equilibrium. Clearly, any living system is active, using stored energy and functioning far away from any thermal equilibrium state.
\newpage

So, is every active system alive? Clearly not. But, what does it take to go from active matter to a living system? The first step involves the predominance of regulatory interactions~\cite{moat2002microbial,schmidt1997animal} . A comparison of laboratory preparations involving active biopolymers~\cite{weitz}  and the actual situation that prevails in the cytoskeleton of a living cell makes the point. In the latter, there are many dozens of proteins that regulate all aspects of the polymer chemistry and couple reactions to cellular conditions. Thereby, actin polymerization is restricted to the front of a moving cell~\cite{weiner}, microtubules attach in a highly controlled manner to segregating chromosomes~\cite{ross}, and intermediate filaments such as Vimentin arrange themselves geometrically to cushion against nuclear deformation~\cite{vimentin}. Active matter physics is slaved to needed functionality, being necessary but not sufficient. Coupling active matter to controlling regulators leads to a new and often qualitatively different class of objects that we will refer to as ``smart matter"~\cite{savoie2019robot,ozkan2021collective}.\\

The operating principles behind regulatory interactions concern information flow, by which we mean that details regarding the external world are used to modulate active behavior. As the relevant environmental inputs vary, the active system responds by realigning its dynamics accordingly; the more complex and often, the more energy-intensive, a regulatory system is, the better it can create useful correlations between the environment and active matter behavior. The active matter can now behave intelligently. And, living systems are clearly very smart in ways we are only beginning to understand. Biology over the past half century has revealed an astounding complexity in the control of all important processes. One can assume with confidence that every step of a biological process will be regulated in multiple ways and over multiple timescales. One can also assume with confidence that individual components of the underlying active matter will evolve to become more flexible in their ability to respond to regulatory input - a case in point is the fascinating story of the evolution of the mammalian synapse~\cite{grant}, where the sophistication of the molecular machinery has grown even as the neural systems it serves have  become larger. Perhaps these facts will eventually validate the musings of Schr\"odinger in his famous ``What is Life” text~\cite{schrodinger1992life} on the need for coming up with new concepts of physics to accommodate the workings of smart matter, workings that just do not occur naturally in the abiotic world.\\

Of course, the boundary between active matter versus smart matter can sometimes be rather fuzzy. Let us take for example the automobile traffic in a big city freeway system. Until quite recently, this could serve as an obvious example of an active matter system that had no overarching regulatory dynamics shaping the local interactions of the active motorists. Even though individual motorists clearly want to minimize their travel time and avoid collisions, there was no goal driving the dynamics of the system as a whole. But, an argument can be made that the advent of tools such as Waze and Google Maps has indeed provided the regulatory feedback missing in the active matter paradigm. Now, motorists do modulate their interactions and decisions based on freeway conditions on a variety of length scales, and the network as a whole does attempt to optimize transportation functionality. In fact, the city of Boston has an ongoing partnership with Waze in which data is fed to the city's traffic management center in order to adjust traffic signals, explicitly meant to optimize transportation efficiency. As surprising as that may seem to those of us who live there, Boston traffic may be becoming intelligent. Parenthetically, this example serves to emphasize that we do not restrict the use of the word "matter" to tangible physical "stuff"; matter to us just means a substrate upon which the actions of interactive components takes place.  This generalization might become particularly important in the future, when we are forced to confront the question of living artificial intelligences.\\

Should we consider any smart matter system to be alive? The question of what it means to be alive is fraught with millennia of philosophical debate that is hard to place in a physical science context. Nonetheless, living systems might require something beyond being simply smart matter possessing efficient information flow governing active medium response. Let us turn to another example in which smart management of environmental interactions is critical for function. The engineering community has devoted considerable effort to building multi-legged robots that can navigate effectively over difficult terrain~\cite{guizzo2019leaps,saranli2001rhex,raibert1986legged}. To do this, the robots are typically equipped with a complex suite of sensors that provide input for controlling sophisticated actuators and/or leveraging embodied (aka mechanical) "intelligence"~\cite{pfeifer2006body} in a by now familiar smart matter pattern. However, despite the impressively life-like agility and performance increasingly being displayed by such devices, it is unlikely that many investigators would consider such a  robot to be alive, in the same sense that an insect navigating the same terrain would be. Why? We can extend the question by imagining that we endow the robot with a battery sensor such that when it detects that its power level is getting low, it stops what it is doing and goes off in search of "electric food" from the nearest charging station. Is the robot now closer to being a living system? What if there is in addition a sensor detecting potentially hazardous weather conditions and the robot ``knows" to seek shelter. What if the robot can decide, based on the rate of progress it is making on an assigned task, that it needs more copies of itself and can arrange to have that happen by ordering from a factory with which it is in contact. Thus, the question posed to physics of living systems researchers is whether the difference between living and smart is just one of degree of systems integration and semantics, or alternatively involves a true phase transition leading to new capabilities in a discontinuous manner. There is no real hint at present as to what might cause such a transition, or if the transition/bifurcation concept is even relevant. The same question arises of course in the microscopic realm, concerning the ancient origin of life and modern attempts to artificially synthesize living cells and to characterize what a minimum cell must consist of~\cite{minimal}. And, this is becoming ever more critical, as advances in astronomy, such as the exoplanet revolution, have brought to the fore questions of how best to search for indications of life elsewhere in the cosmos.\\

We note that even without discussing if such systems are ``alive" we can utilize smart matter systems as models to discover principles by which living systems achieve robust function~\cite{aina2022toward}. Indeed, physicists have a long history of being interested in deep questions but not letting these get in the way of making tangible progress; any history of quantum theory will clearly attest to this useful duality. So, there is much work to done in figuring out how to best couple active matter to smart controllers to enable the accomplishment of various tasks. In this regard, we will need to work directly on all manner of living systems and with all manner of biologists. In some parts of this endeavor, there is a need for mutual re-education. While physicists are most comfortable with the active matter paradigm, modern biological research often stresses the regulatory aspects of living systems at the expense of working back down to physical processes that interact with the environment. Many papers focus almost exclusively on gene expression (aka transcriptomics) and protein abundance as the ultimate in defining cell states and cell physiology. This perspective is becoming even more entrenched as spatial transcriptomics~\cite{spatial} and technologies such as tissue CyTOF~\cite{cytof} begin to define tissues and organs solely in terms of omics profiles. It imagines real-world action as something which can automatically flow from the information processing level - there's always ``a gene for that". An example of the problem arose quite a long time ago in a paper~\cite{palsson} claiming to have found {\em the} gene responsible for creating the chemical wave field responsible for guiding the motion of hundreds of thousands of Dictyostelium amoebae towards aggregation centers, as part of their survival strategy in the face of starvation. A gene and its single protein product cannot make a millimeter scale chemical pattern; instead a gene can help control a physiochemical system of interacting components that are capable of doing the necessary spadework. \\

We therefore posit that the physics of living systems can benefit from a synergistic merging of these two insufficient worlds views; creation of the field of ``smart matter" can provide researchers a way to frame such an integration and develop new models of living systems. In our opinion, this merger is essential. Ignoring the constraints placed on living systems by the need to get the molecules, cells, tissues and organs to actually accomplish the needed tasks will miss essential constraints on behavior. Assuming that active matter systems are all we need to focus on as we move forward dismisses out of hand many of the performance aspects of systems that allow us to consider them living. Molecular and cellular biologists should realize genes are not magic wands that can wish physical effects into existence.  Active and soft matter physicists need to take to heart a quote from Alan Turing regarding patterning of the zebra, that ``..the stripes are easy, but what about the horse part?".

\begin{acknowledgments}
H.L. thanks the NSF Physics Frontier Center program for support. D.I.G. thanks numerous colleagues for helpful discussion over the years and the NSF Physics of Living Systems program, the Army Research Office MURI on algorithmic matter, and the Dunn Family Professorship.
\end{acknowledgments}

\nocite{*}

\bibliography{apssamp}

\end{document}